# A Comparative Study of Cohesive Zone Models for Predicting Delamination Behaviors of Arterial Wall

Ting Miao[1], Liqiong Tian[2], Xiaochang Leng[3], Zhamgmu Miao[4], Chengjun Xu*

## Abstract

Arterial tissue delamination, manifested as the failure between arterial layers, is a critical process in the rupture of atherosclerotic plaque, leading to potential life-threatening clinical consequences. Numerous models have been used to characterize the arterial tissue delamination. Few has investigated the effect of Cohesive Zone Model (CZM) shapes on predicting the delamination behavior of the arterial wall. In this study, four types of cohesive zone models (triangular, trapezoidal, linear-exponential and exponential-linear) were investigated to compare their predictability of the arterial wall failure. The Holzapfel-Gasser-Ogden (HGO) model was adopted for modelling the mechanical behavior of the aortic bulk material. The simulation results using CZM on the aortic media delamination were also compared with the results on mouse plaque delamination and human fibrous cap delamination. The results show that: 1) the simulation results based on the four shapes of CZMs match well with the experimental results; 2) the triangular and exponential-linear CZMs are in good agreement with the experimental force-displacement curves of mouse plaque delamination; 3) considering the viscoelastic effect of the arterial tissue, the triangular and exponential-linear CZMs match well with the experimental force-displacement curves of human fibrous cap delamination. Thus, triangular and exponential-linear CZMs can capture the arterial tissue failure response well.

1School of Transportation, Wuhan University of Technology, Wuhan, China, maioting1615@hotmail.com

2Jiangxi Electric Power Design Institute, Nanchang, China, tianliqiong123@sohu.com

3Institute of Engineering Mechanics, Nanchang University, Nanchang, China, lengxc1984@163.com

4 School of Transportation, Wuhan University of Technology, Wuhan, China, zmmiao@whut.edu.cn

**\* Corresponding author**:

Chengjun Xu, School of Logistics Engineering, Wuhan University of Technology, 1178 Heping ave, Youjiatou campus, Wuchang District, Wuhan 430063, China.

Email: xuchengjun@whut.edu.cn

# 1. Introduction

Each year in the United States, 1.1 million people suffer from myocardial infarction with 40% fatality rate[1]. The arterial tissue failure in terms of arterial wall delamination and separation is the main reason of myocardial infarction. The arterial wall is composed of three layers: intima, media and adventitia. Mechanical properties of each layer of arterial wall depend on its composition, microstructure and mechanical properties of the major load bearing constituents. Hypertension or trauma may induce geometrical mismatch and interlayered shear stress concentration between layers[2,3]. Moreover, intra-layer dissection of the arterial wall in terms of split and cleavage usually occurs in the media[4,5]. With the dissections inside the medial layer, blood will bulge the delaminated media layer, reducing the lumen of the blood vessel and compromising the strength of the arterial wall. Therefore, there is a pressing need to investigate the mechanism of the dissection at the medial layer.

Numerous experimental studies focused on the evaluation of arterial medial layer dissection. Sommer et.al[5] investigated the dissection of human abdominal medial layers via peeling tests and quantified the dissection energy. Tong et.al[6] studied the dissection properties and mechanical strength of the carotid artery and found that interfacial ruptures mainly occurred at the media. Thus, the study on the delamination behaviors of arterial wall, especially in the medial layer is of great importance.

Cohesive zone mode (CZM) approach has been employed to model the medial layer dissection. Gasser et.al[7] implemented the CZM to model the arterial layer dissection and the obtained gap-displacement vs. load/width curves were in good agreement with the experimental measurements[5]. Ferrara et.al[8] utilized an anisotropic cohesive zone model in the numerical simulation to investigate the inter-lamellar tear in the media[5] and captured the main response of arterial tissue failure.

It has been shown that the triangular CZM and the exponential CZM are usually used in the simulations, playing an important role in the traction-separation behavior at the delamination front during the tearing propagation, especially for metal and composite materials[9-11]. The results in these works demonstrate that the ductile or softening behavior of material can affect the strength predictions of the CZM. Moreover, the mechanical response of arterial tissue is largely governed by the orientation and mechanical properties of collagen fibers during the crack initiation and propagation, giving an exponential increase in the stress-strain relationships of arterial wall. However, few studies have investigated the effect of CZM shapes on the delamination behavior of arterial tissue.

This study focuses on the comparison of four different shapes (triangular, trapezoidal, linear-exponential and exponential-linear) of cohesive zone mode in modeling the arterial medial peeling. Meanwhile, as a continuation of our previous studies, we used four types of CZM laws to model the delamination of atherosclerotic plaque and fibrous cap[12,13]. The findings of this study will help us study the influence of cohesive law shapes on the tearing propagation behavior in the medial layer of arterial tissue. To be more specific, simulating the arterial tissue dissection to predict the delamination behaviors of arterial wall promotes the development of effective techniques for performing the treatment and intervention.

## 2. Material and Methods

In this study, the Holzapfel-Gasser-Ogden (HGO) model[14] was used to characterize the anisotropic hyperelastic mechanical behavior of arterial tissue. Four types of cohesive zone models (triangular, trapezoidal, linear-exponential and exponential-linear) were used for modeling the delamination behavior along the medial interface in the arterial wall.

### 2.1 Bulk Material Model: Holzapfel-Gasser-Ogden (HGO) model

The free energy potential $\Psi$, which defines per unit reference volume of arterial material in a decoupled form, is expressed as:

$$\Psi = \Psi_{vol} + \bar{\Psi}. \tag{1}$$

The volumetric part, $\Psi_{vol}$ is given by the work[14] as:

$$\Psi_{vol} = \frac{1}{D}\left(\frac{J^2-1}{2} - \ln J\right), \tag{2}$$

where $\frac{1}{D}$ is analogous to the bulk modulus of the material. The isochoric potential is

$$\bar{\Psi} = \bar{\Psi}_g + \bar{\Psi}_f, \tag{3}$$

which is composed by two parts. One is used to describe the mechanical response of matrix,

$$\bar{\Psi}_g = \frac{\mu}{2}(\bar{I}_1 - 3). \tag{4}$$

Another is implemented to represent the mechanical behavior of two families of collagen fibers,

$$\bar{\Psi}_f(\bar{C}, H_1, H_2) = \frac{k_1}{2k_2}\left[e^{k_2[\kappa \bar{I}_1 + (1-3\kappa)\bar{I}_{41} - 1]^2} - 1\right] + \frac{k_1}{2k_2}\left[e^{k_2[\kappa \bar{I}_1 + (1-3\kappa)\bar{I}_{42} - 1]^2} - 1\right], \tag{5}$$

where $\bar{I}_1 = tr(\bar{C})$ denotes the first invariant of $\bar{C}$, and μ is the neo-Hookean parameter, which characterizes the shear modulus of the material without fibers; $\bar{I}_{41} = \boldsymbol{a}_{01} \cdot \bar{\boldsymbol{C}} \boldsymbol{a}_{01}$ and $\bar{I}_{42} = \boldsymbol{a}_{02} \cdot \bar{\boldsymbol{C}} \boldsymbol{a}_{02}$ are tensor invariants equal to the square of the stretch in the direction of $[\boldsymbol{a}_{01}] = [cos\gamma, \ sin\gamma, 0]$ and $[\boldsymbol{a}_{02}] = [cos\gamma, \ -sin\gamma, 0]$, respectively. Note that constitutive parameter $k_1$ is related to the relative stiffness of fibers, which is determined from mechanical tests of tissue; $k_2$ is dimensionless stiffness. The parameter $\kappa$ is the dispersion parameter, which characterizes the dispersion of the two families of fibers along the two mean distributed directions, and $0 \leq \kappa \leq 1/3$; the collagen fibers of one family of fibers are parallel to each other when $\kappa = 0$, whereas the fibers distribute isotropically when $\kappa = 1/3$. $\gamma$ represents the angle between the mean fiber orientation of one family of fibers and the circumferential direction of the aorta.

**2.2 Interfacial Cohesive Zone Model**

Four types of CZMs are implemented in this study through a user-defined subroutine UEL in ABAQUS[15]. CZM characterizes the relationship between the cohesive tractions of an interface and the separations across the interface-the 'effective' opening displacement $\delta$ which is given by

$$\delta = \sqrt{\lambda^2 \delta_s^2 + \delta_n^2}, \tag{6}$$

where $\lambda$ is a scalar parameter being introduced to assign different weights to the opening displacement $\delta_n$ and sliding displacement $\delta_s$. The parameter $\delta_s$ is denoted by

$$\delta_s = \sqrt{\delta_{s1}^2 + \delta_{s2}^2}, \tag{7}$$

where $\delta_{s1}$ and $\delta_{s2}$ denote components of the two directions of the sliding displacement $\delta_s$ across the cohesive surfaces.

According to the theory of CZM, the delamination failure of arterial layers involves three steps. The first step is damage initiation. When the 'effective' opening displacement reaches $\delta_0$ (for triangular, linear-exponential and exponential-linear CZMs, Fig.1) or $\delta_1$ (for trapezoidal CZM, Fig.1), the cohesive elements start to degrade. Meanwhile, the effective traction t (Cauchy stress) reaches the cohesive strength $\sigma_c$ of the interface. The expressions of effective traction $t$ and the first-order partial derivative of the effective traction $t$ with respective to effective opening displacement $\frac{\partial t}{\partial \delta}$ are shown in the *Appendix A*. The second step is damage evolution, during which damage accumulation occurs in the cohesive element. The damage variable $d$ (expressions for four types of CZMs are shown in *Appendix A*) is introduced in order to quantitatively evaluate the damage. Evidently, $d$ ranges from 0 to 1, which correspond to no damage of the cohesive surface and a fully separation of the cohesive surface, respectively. Fig. 1 illustrates the loading and unloading processes of the cohesive traction-separation law. The damage accumulates when the effective traction $t$ exceeds $\sigma_c$, where at the point A in Fig. 1. The traction-separation relation will go along line AO and BO when unloading at point A and B, respectively, since a permanent damage occurs on the cohesive surface at points A and B. On the contrary, the traction-separation relation will go along the original curve when no damage occurs on the cohesive surface. The third step is the completed failure of the cohesive element after the effective displacement reaches a critical effective separation value $\delta_c$ (as shown in Fig. 1.). Meanwhile the critical energy release rate $G_c$ is attaned.

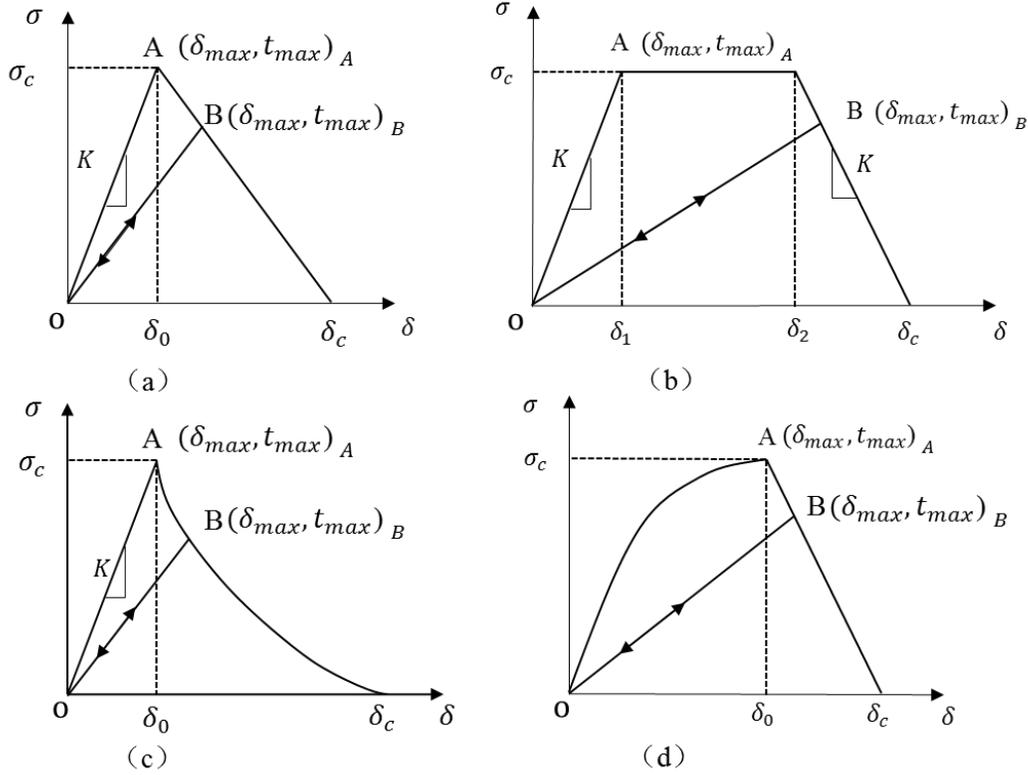

**Fig. 1.** Four types of irreversible cohesive model denotes with effective traction and effective displacement; (a) triangular CZM; (b) trapezoidal CZM; (c) linear-exponential CZM; (d) exponential-linear CZM.

### 3   Numerical implementations

In this section, in order to predict the delamination behaviors of the arterial tissue, a part of the arterial wall is utilized to do experiment and simulation. The sampling method for the experiment is shown in Fig.2. The geometrical model is reconstructed based on the experiments of peeling test of human aortic media[5], as shown in the red box of Fig.2c. The geometrical dimensions of the arterial wall are 4mm of length, 1.2mm of width and 0.05mm of thickness. The front and back surfaces of the strip (Fig. 2c, in the red line box) are fixed along z direction for the sake of computational efficiency. The delamination behavior is modeled as plane strain deformation.

To evaluate the performance of the arterial wall dissection, the finite element (FE) simulations are implemented using ABAQUS[15-17], associated with a user-defined subroutine modeling the delamination of the cohesive interface in the arterial layer. The arterial layer is meshed with 1920 eight-node brick elements (C3D8H), while the cohesive interface is meshed with 72 zero thickness eight-node 3D user-defined elements. A layer of cohesive elements is placed along the delamination path starting from the initial peeling front to the end of arterial strip. The size of the elements for the FE model is 0.05mm based on the parametric analysis. The implementation of CZM in the simulation of interfacial material separation events has high computational efficiency when the peeling path is predefined. In the current study, the peeling path is obtained from the experimental observations.

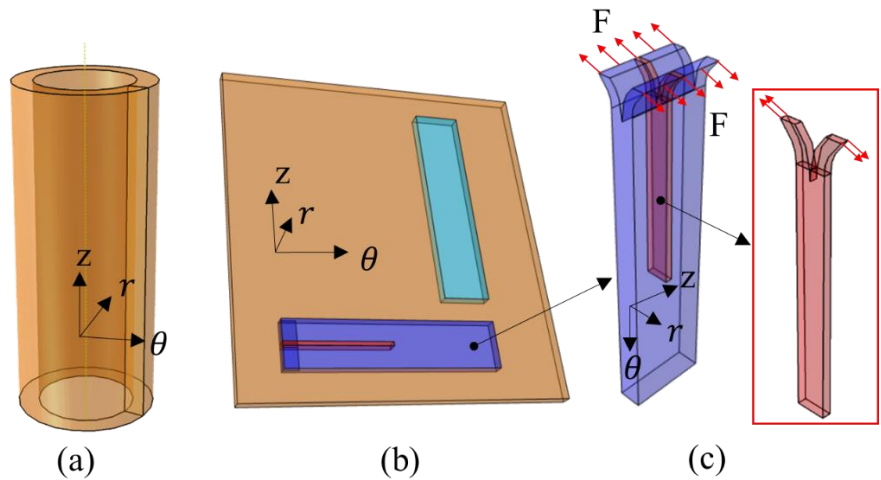

**Fig. 2** A schematic of arterial layer peeling test setup: (a) a part of aorta; (b) a flat of arterial layer after cutting from (a); (c) an arterial strip which was peeled at one end and the geometric model of the peeling test for the simulation (in the red line box).

The parameter values for the HGO and CZM models were taken from references[7, 8, 18], as shown in Table 1. The value of the interfacial stiffness K was assigned as the same one from reference[12]. With regard to the simulations of mouse plaque delamination and human fibrous cap delamination processes, the material parameter values were adopted from the references [12, 13], including viscoelastic material model parameters for human arterial tissue[13], Meanwhile, the experimental setup and simulation modeling details are all presented in the references [12, 13].

**Table 1.** Material parameter values for arterial layer and CZM parameter values for cohesive layer

| HGO model | $\mu$ [MPa] | $k_1$ [MPa] | $k_2$ | $\gamma$ [Deg] | $\kappa$ |
|---|---|---|---|---|---|
| | 0.0162 | 0.0981 | 10 | 5 | 0 |
| CZM | $G_c$ [N/mm] | $\sigma_c$ [MPa] | $K$ [N/$mm^3$] | $\lambda$ | |
| | 0.049 | 0.14 | 1e3 | 10 | |

## 4  Results

The parameters used for those simulations of artrial layer peeling test (as shown in Fig.2) are set with the same values as shown in Table 1. The pulling forces divided by the width of specimen vs. the load-point displacement are shown in Fig. 3. The pulling forces per unit of width is between 25.5 mN/mm and 28.5 mN/mm, which are in the domain of the experimental results (23-35 mN/mm).

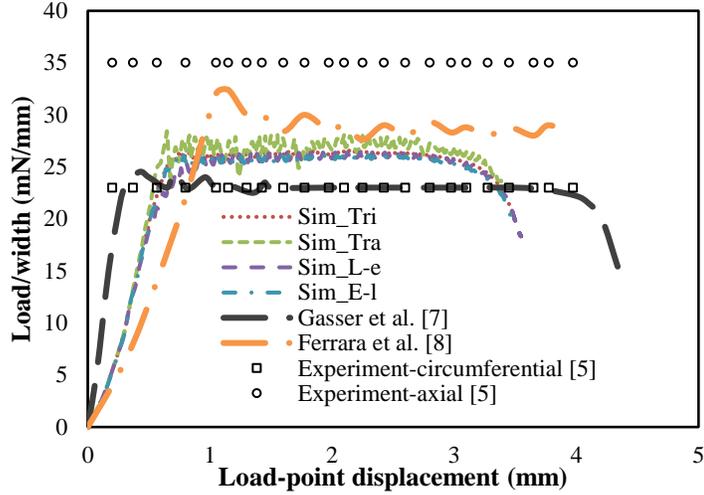

**Fig. 3.** Comparison of the load/width-load-point displacement curves between simulation and average experimental results.

The deformed configurations of the aortic medial delamination are shown in Fig.4a. The contour levels shown in the figures represent the normal component of the Cauchy stress to the delamination plane. In term of a cohesive element in the middle of the cohesive path, the load-point displacement relationships were presented in Fig.4b.

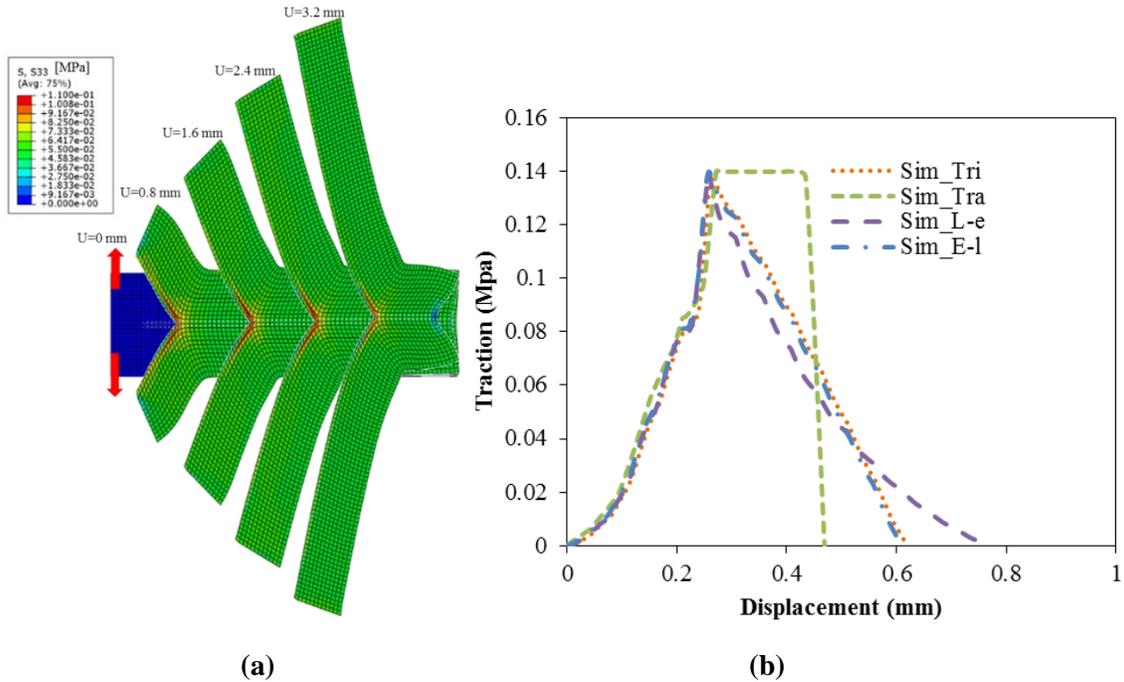

**(a)**              **(b)**

**Fig. 4.** (a) Tensile stress contour of arterial wall at five levels of the aortic media peeling; (b) The load-point displacement relationship of a cohesive element in the middle of the cohesive path.

The load-displacement curves are shown in Fig.5a and Fig.6a. The traction-relative displacement relationships of a cohesive element in the middle of the cohesive path for the plaque and fibrous cap are shown in Fig. 5b and Fig. 6b. The penalty stiffness of the CZM for the simulation of plaque and fibrous cap delamination are $10^4$ N/$mm^3$ and 10 N/$mm^3$, respectively.

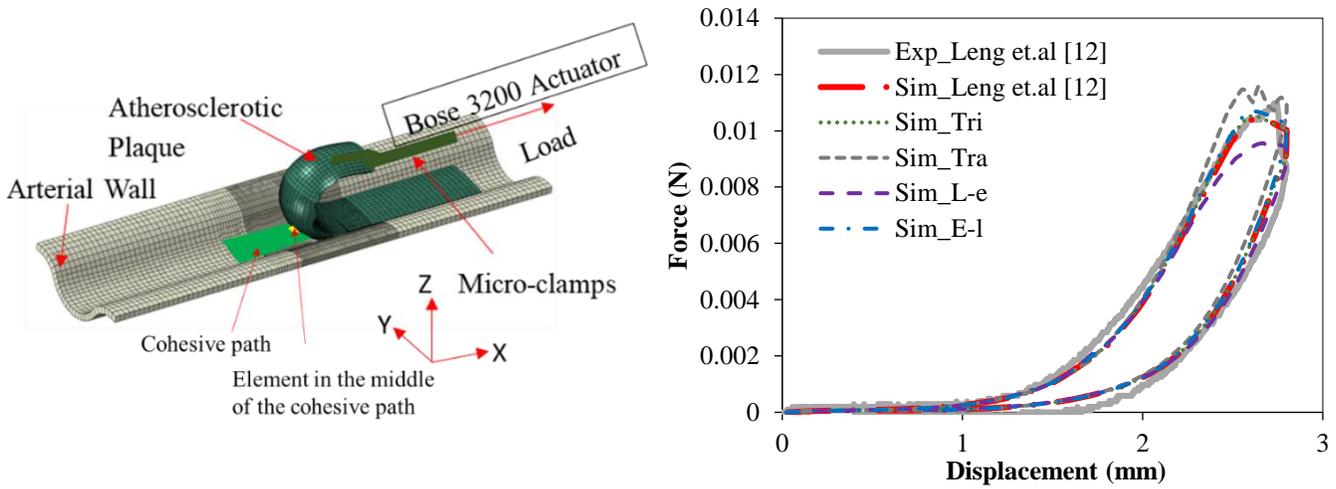

(a)

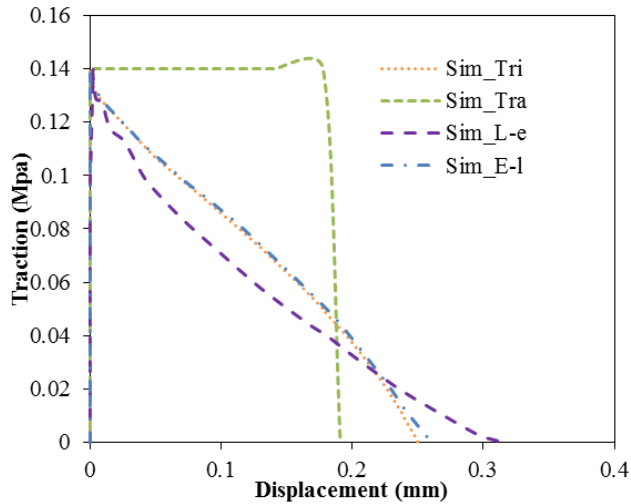

(b)

**Fig. 5.** The simulation-predicted load-displacement curves of loading-delamination-unloading cycles are compared with the experimental measured curves: (a) mouse atherosclerotic plaque delamination and (b) the traction of relative displacement curve of a cohesive element in the middle of the cohesive path.

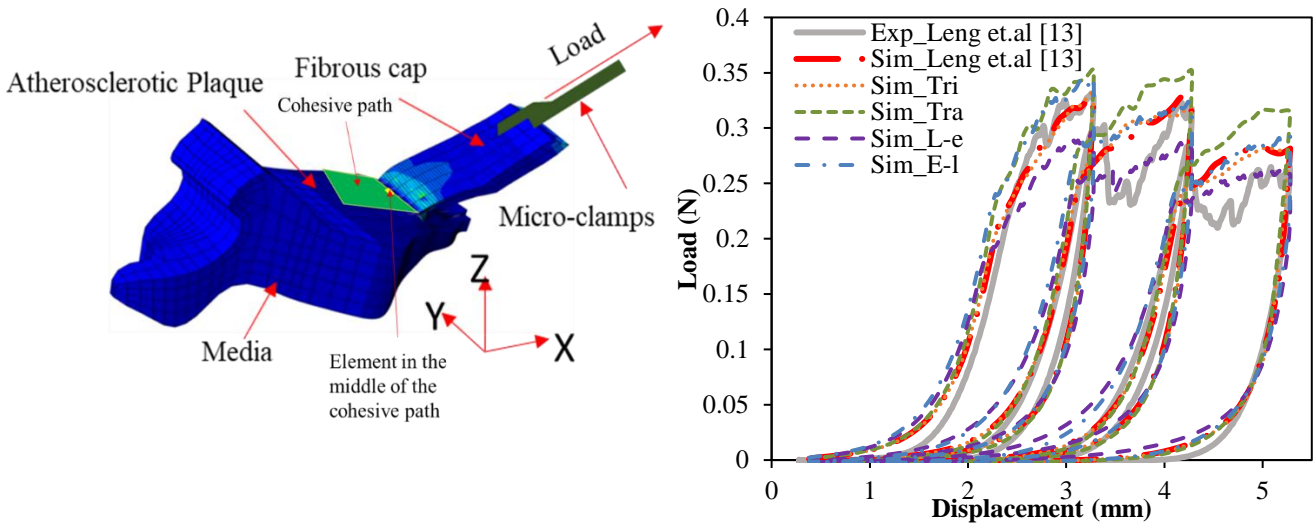

(a)

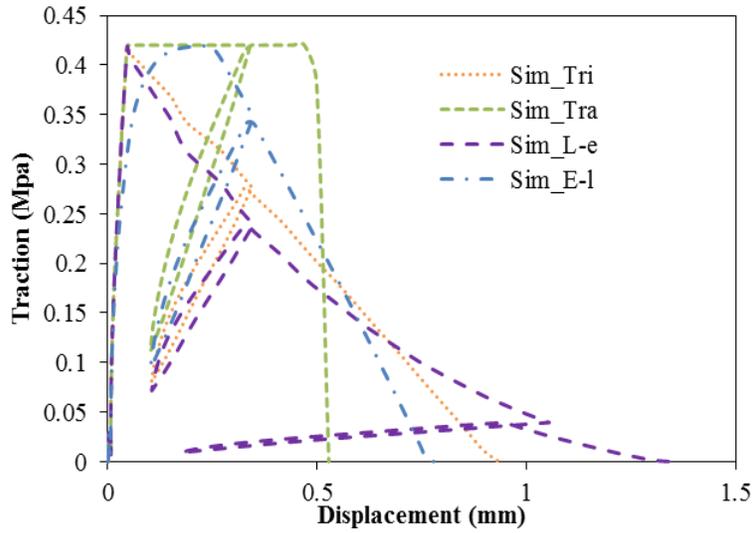

(b)

**Fig. 6.** The simulation-predicted load-displacement curves of loading-delamination-unloading cycles are compared with the experimental measured curves: (a) human fibrous cap delamination and (b) the traction of relative displacement curve of a cohesive element in the middle of the cohesive path.

## 5. Discussion

The aim of this study is to compare four different shapes (triangular, trapezoidal, linear-exponential and exponential-linear) of cohesive zone mode in modeling the arterial medial peeling (as shown in Fig.3 and Fig.4). Moreover, four types of CZM laws were used to model the delamination of atherosclerotic plaque (Fig.5) and fibrous cap[12, 13] (Fig.6).

The comparison of four shapes of CZM laws for modeling aortic medial delamination was described in terms of the load/width vs. load-point displacement, tensile stress contour and traction-displacement curves. This study simulated the aortic media delamination experiments in Sommer et al.[5] by four types of CZM, and further compares the results with simulations in Gasser et al.[7] and Ferrara et al.[8]. In Fig.3, some data recorded in this study are within the range of published results. The pulling forces per unit of width is between 25.5 mN/mm and 28.5 mN/mm, which are in good agreement with the numerical results by Gasser et al. (23 mN/mm)[7] and Ferrara et al. (28.8 mN/mm)[8], which also based on experiments in Sommer et al.[5]. It is observed that the shape of the softening stage is almost the same with the shape of corresponding CZM, as shown in Fig.4b. Comparing with other shapes of CZMs, the minimum values of the traction (linear-exponential CZM) (Fig.4b) reveal that the reaction forces (Fig. 3) during the medial delamination are smaller than that of other types of CZMs.

Besides the results comparison on the experiment of the aortic media delamination, the simulation of mouse plaque delamination[12] and human fibrous cap delamination[13] were carried out to compare with the simulation results obtained by using four types of CZM. A mesh dependency study of mouse plaque and human fibrous cap delamination has been conducted by previous work[12,13,19] in which the selected mesh refinement is enough to obtain the convergence solutions. For the two delamination simulations, the magnitude of predicted maximal force during delamination by linear-exponential CZM is lower by nearly 10%, while the magnitude of trapezoidal CZM is higher by nearly 4% than what reported in[13,20]. The traction of the elastic stage of the CZM for plaque delamination rapidly increases to the maximum value. For the fibrous cap delamination, the traction of the elastic stage of the CZM is similar to the shape of CZM (Fig. 6b). It shows that the shape of the softening stage is almost the same as the shape of corresponding CZM. The minimum values of the traction (linear-exponential CZM) compared to other shapes of CZMs reveals that the reaction forces (Fig. 5a, 6a) during the plaque and the fibrous cap delamination are less than the other types of CZMs. Unloading and loading processes of the element under unloading and loading cycles are shown in Fig. 6b, indicating the softening of the element experiences a long loading displacement.

The shapes of the traction-displacement curves of an element at the front of the crack are in good agreement with the shape of CZM laws used to simulate the aortic delamination process, showing that the extent of damage evolution is the largest for linear-exponential CZM, followed by triangular, exponential-linear and trapezoidal CZM laws. Moreover, this respective order does not represent the order of the predicted load/width values. The trapezoidal law shows higher values of transmission of the load/width at the initial stage of damage, because the maximum traction kept constant under the longest displacement load during the damage evolution process compared with other shape of CZM laws. The triangular CZM can obtain the predicted load/width or load with an acceptable accuracy, as well as it is the easiest CZM law to use in term of simulation of arterial tissue failure. The iterative process and the time for acquiring the solutions are influenced by the shape of CZM laws. The degree of difficulty to get convergence solutions in the iterative process is the highest for trapezoidal, followed by linear-exponential, exponential-linear and triangular CZMs. At the stage when a cohesive element is totally failed, the release area of the bulk material will induce a contraction, which indicates that the oscillation for the predicted load by using trapezoidal CZM is more obvious than other CZM laws.

The elastic deformation of the cohesive element is governed by the mechanical properties of its structural constituents and the interfacial strength of material (e.g. triangular, trapezoidal CZM…). The nonlinearity of the traction-displacement relationship of cohesive element is largely attributed to the gradual recruitment of the load-bearing collagen fibers with increasing pressures[21,22,23]. The softening of the traction-displacement relationship is attributed to the softening behavior of the medial interfacial strength, plaque or fibrous cap[24,25,26]. Hence, these cohesive zone models are able to characterize the mechanical response of cohesive element during the dissection propagation.

The limitations of this study are associated with the introduced assumptions. The variation of collagen fiber orientation is partially considered for by the fiber-reinforced constitutive model. However, it has been shown a continuous distribution of fiber orientations[27, 28]. Moreover, the proposed layered constitutive model assumes that the arterial wall composition and structure are uniform in the radial direction. More detailed assessment of arterial wall composition and structural may obtain more accurate numerical predictions of arterial tissue failure.

## 6. Conclusions

In this paper, four types of CZMs have been compared to investigate the effect of CZM shapes on the failure strength predictions by modeling aortic media peeling, mouse plaque delamination and human fibrous cap delamination events. The main observations have been pointed out as follows. First, the numerical predictions which based on the triangular, trapezoidal, linear-exponential and exponential-linear CZMs match well with the experimental measurements of aortic media peeling process. Secondly, the triangular and exponential-linear CZMs match well with the experimental force-displacement curves, while the trapezoidal CZM over predicts and linear-exponential CZM under predicts the force of mouse plaque delamination. Thirdly, considering the viscoelastic effect of the arterial tissue, triangular and exponential-linear CZMs match well with the experimental force-displacement curves, whilst the trapezoidal CZM over predicts and linear-exponential CZM under predicts the force of human fibrous cap delamination.

**Appendix**[29]

**Triangular CZM**

A triangular form of energy release rate for the cohesive surface can be expressed as

$$G_c = \frac{1}{2}\sigma_c \delta_c. \tag{A1}$$

The displacement jump for the damage initiation and complete damage of the cohesive element denote by

$$\delta_0 = \frac{\sigma_c}{K}, \tag{A2}$$

$$\delta_c = \frac{2G_c}{\sigma_c}, \tag{A3}$$

where $K$ is a penalty stiffness.

The effective traction can be expressed as

$$t = \begin{cases} \frac{\delta}{\delta_0}\sigma_c &, \delta \leq \delta_0 \\ \sigma_c\left(\frac{\delta_c - \delta}{\delta_c - \delta_0}\right) = \sigma_c(1-d) &, \delta_0 \leq \delta < \delta_c \\ 0 &, \delta_c \leq \delta. \end{cases} \tag{A4}$$

The first-order partial derivative of the effective traction $t$ with respective to effective opening displacement is given by

$$\frac{\partial t}{\partial \delta} = \begin{cases} \frac{\sigma_c}{\delta_0} &, \delta \leq \delta_0 \\ \frac{-\sigma_c}{\delta_c - \delta_0} = \frac{-\sigma_c d}{\delta - \delta_0} &, \delta_0 \leq \delta < \delta_c \\ 0 &, \delta_c \leq \delta, \end{cases} \tag{A5}$$

where $d$ is the Damage variable defined to represent the softening of the cohesive element

$$d = \begin{cases} 0 &, \delta \leq \delta_0 \\ \frac{G_c - \left[\sigma_c\left(\frac{\delta_c - \delta}{\delta_c - \delta_0}\right)\delta_c\frac{1}{2}\right]}{G_c} = \frac{\delta - \delta_0}{\delta_c - \delta_0} &, \delta_0 \leq \delta < \delta_c \\ 1 &, \delta_c \leq \delta. \end{cases} \tag{A6}$$

**Trapezoidal CZM**

A trapezoidal form of energy release rate for the cohesive surface can be expressed as [30]

$$G_c = \frac{1}{2}\sigma_c[(\delta_2 - \delta_1) + \delta_c]. \tag{A7}$$

The displacement jump for the damage initiation, traction softening initiation and complete damage of the cohesive element denote by

$$\delta_1 = \frac{\sigma_c}{K}, \tag{A8}$$

$$\delta_2 = \frac{G_c}{\sigma_c}, \tag{A9}$$

$$\delta_c = \delta_1 + \delta_2, \tag{A10}$$

where $K$ is a penalty stiffness.

The effective traction can be expressed as

$$t = \frac{\partial \varphi}{\partial \delta} = \begin{cases} \frac{\delta}{\delta_1}\sigma_c, & \delta \leq \delta_1 \\ \sigma_c, & \delta_1 \leq \delta < \delta_2 \\ \frac{(1-d)(\delta_2 - \delta_1 + \delta_c)}{\delta_c}\sigma_c, & \delta_2 \leq \delta < \delta_c \\ 0, & \delta_c \leq \delta. \end{cases} \tag{A11}$$

The first-order partial derivative of the effective traction t with respective to effective opening displacement is given by

$$\frac{\partial t}{\partial \delta} = \begin{cases} \frac{\sigma_c}{\delta_1}, & \delta \leq \delta_1 \\ 0, & \delta_1 \leq \delta < \delta_2 \\ -\sigma_c \frac{(1-d)(\delta_2 - \delta_1 + \delta_c)}{\delta_c(\delta_c - \delta)}, & \delta_2 \leq \delta < \delta_c \\ 0, & \delta_c \leq \delta \end{cases} \tag{A12}$$

where $d$ is the Damage variable defined to represent the softening of the cohesive element

$$d = \begin{cases} 0, & \delta \leq \delta_1 \\ \frac{(\delta - \delta_1)}{(\delta_2 - \delta_1 + \delta_c)}, & \delta_1 \leq \delta < \delta_2 \\ 1 - \frac{\delta_c(\delta_c - \delta)}{(\delta_c - \delta_2)(\delta_2 - \delta_1 + \delta_c)}, & \delta_2 \leq \delta < \delta_c \\ 1, & \delta_c \leq \delta. \end{cases} \tag{A13}$$

**Linear-exponential CZM**

A linear-exponential form of energy release rate for the cohesive surface can be expressed as

$$G_c = \sigma_c(\delta_c - \delta_0)\left(1 - \frac{1}{1-e^{-\alpha}} + \frac{1}{\alpha}\right), \tag{A14}$$

where $\alpha$ is the shape of exponential law ($\alpha = 2$).

The displacement jump for the damage initiation and complete damage of the cohesive element denote by

$$\delta_0 = \frac{\sigma_c}{K} \tag{A15}$$

$$\delta_c = \frac{G_c}{\sigma_c\left(1 - \frac{1}{1-e^{-\alpha}} + \frac{1}{\alpha}\right)} + \delta_0, \tag{A16}$$

where $K$ is a penalty stiffness.

The effective traction can be expressed as

$$t = \begin{cases} K\delta, & \delta \leq \delta_0 \\ K\delta(1-d), & \delta_0 \leq \delta < \delta_c \\ 0, & \delta_c \leq \delta. \end{cases} \tag{A17}$$

The first-order partial derivative of the effective traction $t$ with respective to effective opening displacement is given by

$$\frac{\partial t}{\partial \delta} = \begin{cases} K, & \delta \leq \delta_0 \\ (1-d)K, & \delta_0 \leq \delta < \delta_c \\ 0, & \delta_c \leq \delta, \end{cases} \tag{A18}$$

where $d$ is the Damage variable defined to represent the softening of the cohesive element

$$d = \begin{cases} 0, & \delta \leq \delta_0 \\ 1 - \frac{\delta_0}{\delta_{max}}\left[1 - \frac{1-\exp\left[-\alpha\left(\frac{\delta_{max}-\delta_0}{\delta_c-\delta_0}\right)\right]}{1-\exp(-\alpha)}\right], & \delta_0 \leq \delta < \delta_c \\ 1, & \delta_c \leq \delta. \end{cases} \tag{A19}$$

**Exponential-linear CZM**

A triangular form of energy release rate for the cohesive surface can be expressed as

$$G_c = \frac{\sigma_c}{K}\left[K\delta_0 - \sigma_c + \sigma_c \exp\left(-\frac{\delta_0 K}{\sigma_c}\right)\right] + \frac{1}{2}\sigma_c(\delta_c - \delta_0). \tag{A20}$$

The displacement jump for the damage initiation and complete damage of the cohesive element denote by

$$\delta_0 = -\frac{In(e)\sigma_c}{K} \tag{A21}$$

$$\delta_c = \frac{2G_c}{\sigma_c} - \frac{2}{K}\left[K\delta_0 - \sigma_c + \sigma_c exp\left(-\frac{\delta_0 K}{\sigma_c}\right)\right] + \delta_0, \tag{A22}$$

where $K$ is a penalty stiffness.

The effective traction can be expressed as

$$t = \begin{cases} \sigma_c\left(1 - \exp\left(-\frac{\delta K}{\sigma_c}\right)\right), & \delta \leq \delta_0 \\ (1-d)(1-e)\sigma_c\frac{\delta}{\delta_0}, & \delta_0 \leq \delta < \delta_c \\ 0, & \delta_c \leq \delta. \end{cases} \tag{A23}$$

The first-order partial derivative of the effective traction $t$ with respective to effective opening displacement is given by

$$\frac{\partial t}{\partial \delta} = \begin{cases} K\exp\left(-\frac{\delta K}{\sigma_c}\right), & \delta \leq \delta_0 \\ (1-d)(1-e)\frac{\sigma_c}{\delta_0}, & \delta_0 \leq \delta < \delta_c \\ 0, & \delta_c \leq \delta, \end{cases} \tag{A24}$$

where $d$ is the Damage variable defined to represent the softening of the cohesive element

$$d = \begin{cases} 0, & \delta \leq \delta_0 \\ 1 - \frac{\delta_0(\delta_c - \delta_{max})}{\delta_{max}(\delta_c - \delta_0)}, & \delta_0 \leq \delta < \delta_c \\ 1, & \delta_c \leq \delta. \end{cases} \tag{A25}$$

**Acknowledgments**

Research reported in this publication was supported by the National Science Foundation (NSF) under Award Number CMMI-1200358. This work was also partially supported by a SPARC Graduate Research Grant to Xiaochang Leng from the Office of the Vice President for Research at the University of South Carolina.